# Photo-strobo-acoustic Imaging at the Microscale by Laser-Induced Ultrasound

Alvarez-Martinez J.U.[1], Polo-Parada Luis[2], Gutierrez-Juarez G[1]., ., Medina-Cazarez O[1]., Castro-Beltran R[1], [1]Universidad de Guanajuato, Campus Leon Division de Ciencias e Ingenierias, [2]Dalton Cardiovascular Research Center - University of Missouri.

March 13, 2023

### Abstract

The combination of microfluidic technology and optical fluids characterization techniques has been recently applied to produce lab-on-a-chip systems. In the present work, bringing together the imaging technique called photoacoustic imaging (PAI) and microfluidic technology were implemented to obtain micro-scale imaging. Laser-induced ultrasound signals were measured from microdroplets produced in a simple T-junction microfluidic system. Single pulse laser images were produced as a result of the combination of the aforementioned techniques, allowing to obtain of geometrical information of the microdroplets and its spatial position.

# 1 Introduction

In the last decade, droplet generation based on microfluidic technology has been widely used in science applications [1]. Emerging interest in microfluidic systems lies in features that allow the analysis of small-volume samples with improved analytical performance [2] compared to macroscale analogs. Microdroplets can occur in different generation regimes depending on the experimental parameters, including squeezing, dripping, and jetting. The formation of drops in each regime is dependent on experimental parameters. For example, in the squeezing regime (0.001 ¡ Ca ¡ 0.1) [3], the geometry of these depends on the flow rates of each of the flows. In a typical experiment, a thousand microdroplets can be produced, then, the ability to analyze the individual microdroplet content represents a technological challenge approach [4]. Nevertheless, recently has been shown the potential to characterize microscale events using the photoacoustic phenomenon (PA), and the photoacoustic imaging (PAI) technique [5]. To generate the PAI-wide options of transducer materials have been reported such as crystals: PMN-PT, LinBo3, or polymer/ceramic piezoelectric: PVDF/PZT, each of the transducers represents pros and cons depending on the material under study. For example, Da-Wei Wu et al showed the capability and efficiency of the PZT to transduce the Laser-Induce ultrasound (LIU) to electrical (PA) signals with a bandwidth of 100MHz. In this sense the conjunction of the microfluidic system and PAI are now considered to produce novel Lab-on-a-chip (LoC) systems. Commonly the PAI are generated from the PA signal and a post-signal treatment software [6].

# 2 Methods

The experiment was carried out considering the T-junction microfluidic system (MS) manufacturing as described in The droplet generation was driven by pumping the fluids flows into the MS through a gravity pump. The MS is placed in a platform where the acoustic transducer sensor (ATS) and the optical fiber are fixed as shown in Figxx. The experiments begin applying 0.5 and 1.8 Pa for the disperse and continuous phases syringe respectively. This corresponds to a flow rate ratio (FRR) $\sim 0.3$ . Due to the droplet generation regime (squeezing) FRR, microchannel dimensions, and geometry the expected droplet lengths are expected to be 400 $\mu m$ The pulsed laser has a 5ns pulse duration with $\lambda = 532$ nm at 10Hz frequency repetition for a fixed energy of 1 mJ for all the experiments.

The experiments begin generating cross-flow of cupric nitrate (CuNO3) and mineral oil in a T-junction microchannel where the high and width were set as $h$ = 50$\mu m$ and $w$ = 200$\mu m$ respectively, generating mono-disperse microdroplets. The aforementioned FRR generates $\sim$ 400 length droplets. The estimated microdroplet volume was calculated to be $\sim$ 100 pico-liters. The optical fiber and the ATS were placed 10$mm$ from the microchannel. The PDMS was used as an impedance coupled with a speed sound of $V_{PDMS} = 1076 \frac{m}{s}$ avoiding an extra interface, thus, enhancing the efficiency of the PAS transmission from the droplet to the ATS.

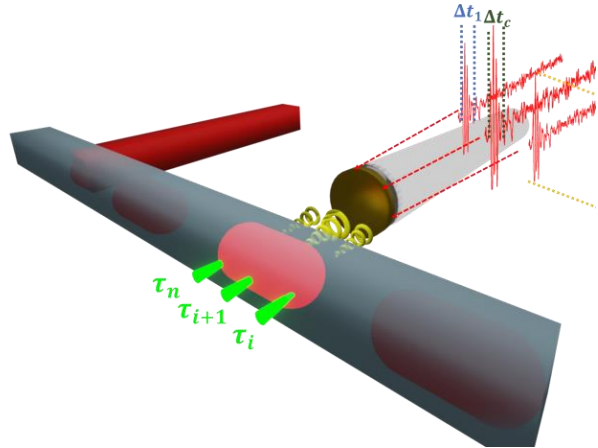

Figure 1: Experimental set-up of an artistic representation. The PA amplitude and pulse time duration are modified accordingly to the incident laser position on the microdroplet excitation zone.

Experimentally it was observed that the PA signal shape is attached to the excitation spatial position on the microdroplet. This fact, allowed us to generate PA imaging with geometrical detail. To do that, each photoacoustic signal measured was subjected to a filtering mathematical method. This method eliminates the ringing and signal noise remaining in the pure photoacoustic response of the fluid.

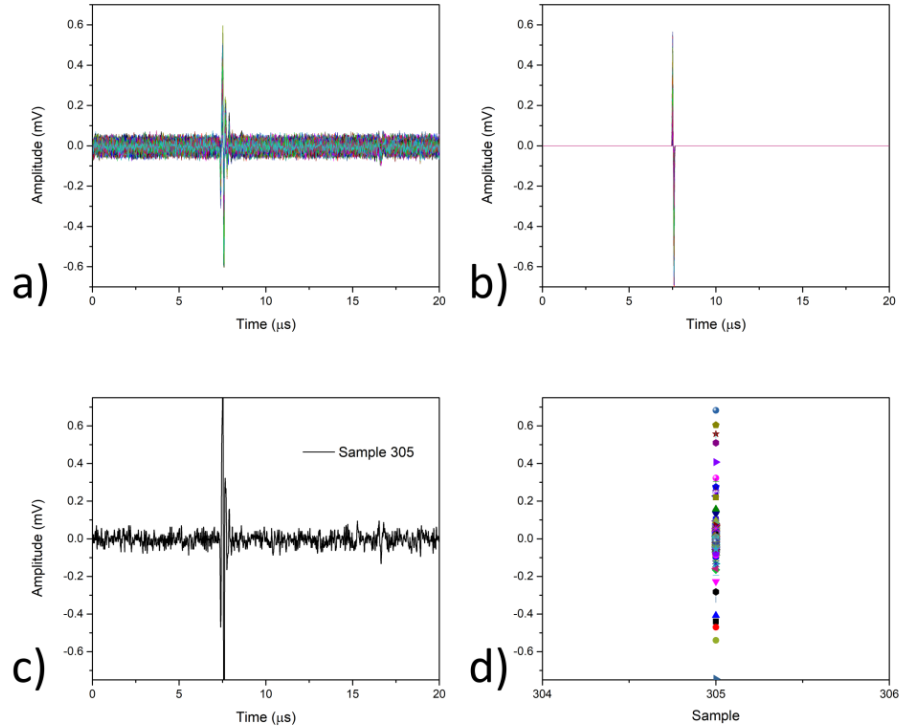

Figure 2: a) Photoacoustic signals from the experimentally acquired data, b) the photoacoustic results from the filtering method, c) a single PA signal from the experiment presented as a vector form, and d) the very same PA signal but presented as a matrix form.

As can be observed in Figure 2, after the filtering method, the photoacoustic signals presented in Figure 2 a) can be cleaned from the ringing effect and the electronic noise, keeping the amplitude and pulse duration of the first PA peaks (positive and negative). Figure 2 c) and d) present the very same signal, for c) the typical PA signal representation vs time duration is plotted and d) the proposed representation where the PA amplitude range is plotted for each experimentally acquired sample number.

This last representation allows us to observe most simply and practically the geometrical influence of the generator object, using the entire PA signal and not only the flight time pulse.

Following this, the entire experimental data can be plotted as a matrix array to observe the differences in the amplitude range for each different geometrical point of laser incidence. Figure 3 shows the plot of the matrix array in the experimental data.

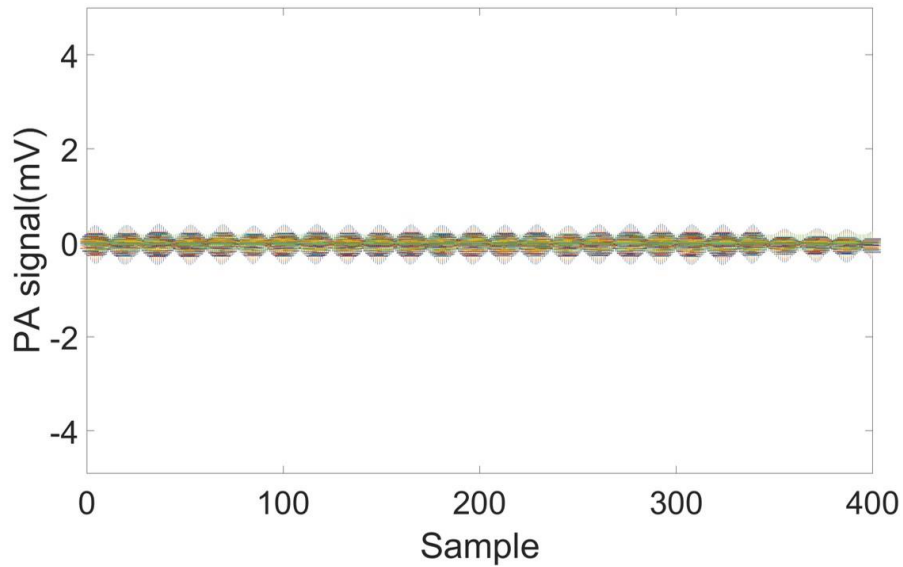

Figure 3: a) Photoacoustic signals from the experimentally acquired data are presented in a matrix form.

Representing the amplitude range in a matrix array gives immediate visual evidence of the objects that generate the PA signals. Moreover, as aforementioned, geometrical objects detail.

## 3   Results

The experimental results and its digital imaging process method allowed us to form time-resolved imaging. The threshold and filtering impact in a highcontrast image, where the geometrical droplet details can be observed. The threshold sets the half maximum amplitude of each acquired signal as a condition to keep the data value or sets the data equal to zero. After, the resulting values are subjected to a Gaussian filter to recover the typical PAs signal improving the image results.

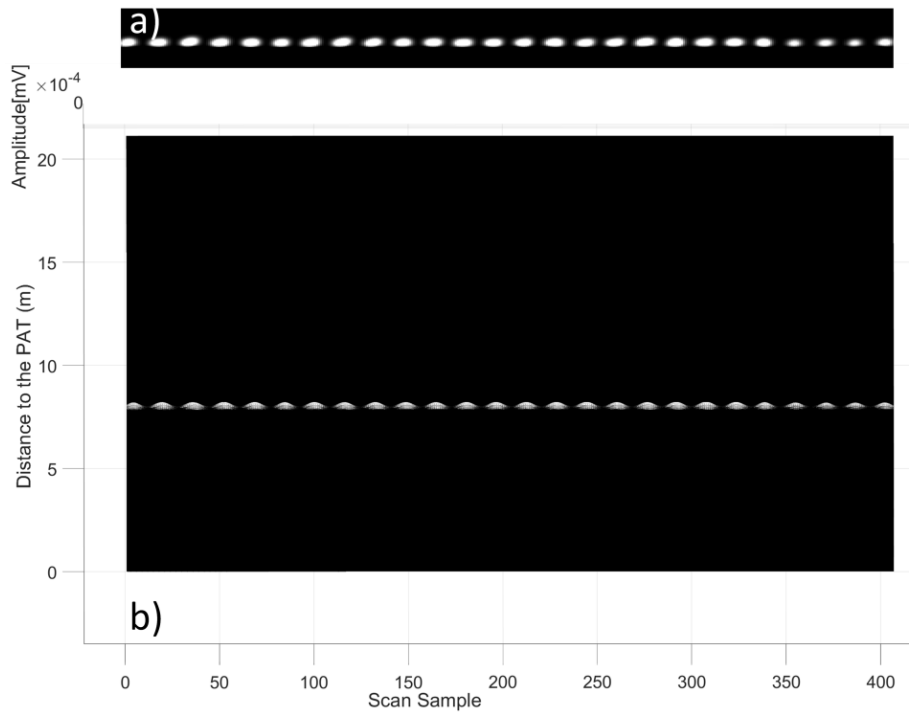

Figure 4: a) and b) respectively shown a 2D and 3D images representation, obtained from the PA imaging method.

Figure 4 a) shows a zoom on the matrix 2D representation for the filtered experimental data, in this, the typical droplet shape can be appreciated, and the time-equidistant separation between the generated droplets, typical in a gravity pump - T-junction droplet generation driven system.